\documentclass[a4paper]{article}

\usepackage{INTERSPEECH2020}
\usepackage{multirow}
\usepackage{graphicx}
\usepackage{float}

\usepackage[ruled,vlined]{algorithm2e}
\usepackage{microtype}
\SetKwInput{KwInput}{Input}
\SetKwInput{KwOutput}{Output}
\usepackage{hyperref}
\usepackage{caption} 
\newcommand\blfootnote[1]{%
  \begingroup
  \renewcommand\thefootnote{}\footnote{#1}%
  \addtocounter{footnote}{-1}%
  \endgroup
}

\title{Data balancing for boosting performance of low-frequency classes in Spoken Language Understanding}
\name{Judith Gaspers$^{*}$, Quynh  Do$^{*}$, Fabian Triefenbach}
\address{Amazon Alexa, Aachen, Germany}
\email{\{gaspers, doquynh, triefen\}@amazon.de}

\begin{document}
\maketitle
\begin{abstract}
Despite the fact that data imbalance is becoming more and more common in real-world Spoken Language Understanding (SLU) applications, it has not been studied extensively in the literature. To the best of our knowledge, this paper presents the first systematic study on handling data imbalance for SLU. In particular, we discuss the application of existing data balancing techniques for SLU and propose a multi-task SLU model for intent classification and slot filling. Aiming to avoid over-fitting, in our model methods for data balancing are leveraged indirectly via an auxiliary task which makes use of a class-balanced batch generator and (possibly) synthetic data. Our results on a real-world dataset indicate that i) %in contrast to common data re-sampling and re-weighting methods, 
our proposed model can boost performance on low frequency intents significantly while avoiding a potential performance decrease on the head intents, ii) synthetic data are beneficial for bootstrapping new intents when realistic data are not available, but iii) once a certain amount of realistic data becomes available, using synthetic data in the auxiliary task only yields better performance than adding them to the primary task training data, and iv) in a joint training scenario, balancing the intent distribution individually improves not only intent classification but also slot filling performance.

\blfootnote{* The first two authors contributed equally to this paper.}
\end{abstract}

% !TEX root = ../coling2020.tex
\section{Introduction}
%Data imbalance is a known problem in machine learning where there is a disproportionate ratio of samples in each class. It poses a challenge for predictive modeling as almost all machine learning classification algorithms is designed around the data balance assumption leading to poor predictive performance, specifically for the minority classes. In real-world applications, data imbalance is a common issue as data is usually collected gradually over application life cycle. Unfortunately, it has not been studied widely in literature especially for modern DNN-based models.
%In this paper we address the data imbalance problem in DNN-based models for Spoken Language Understanding, which is the core of Alexa technology. 
Spoken Language Understanding (SLU) models are essential components in voice-controlled devices, such as Amazon Alexa, Siri and Google Assistant. Typically, SLU addresses the two sub-tasks of intent classification (IC) and slot filling (SF). While the former identifies a speaker's intent, the latter extracts semantic constituents from the natural language query. For example, given the user utterance ``play music by volbeat", the IC subtask should identify \emph{PlayMusic} as intent, while the SF subtask should detect ``volbeat" as \emph{Artist}.
%Consider an example from the ATIS data \cite{conf/slt/TurHH10}:
%\begin{quote}
%    \textbf{city}~~~~~~ [$_{\bf{O}}$~where] %[$_{\bf{O}}$~is] %[$_{\bf{B-airport\_code}}$~MCO].
%\end{quote}
%In this example, NER sub-task should classify ``where'' and ``is'' as ${\bf{O}}$ and MCO as ${\bf{B-airport\_code}}$, the airport code, while IC sub-task should identify \textbf{city} as the speaker's intent.
Recently, mostly DNNs are explored for SLU, which model IC and SF jointly to leverage the interaction between the two subtasks 
%which has been found to improve performance 
(e.g. 
%\newcite{chen19}
\cite{Liu2016AttentionBasedRN}, \cite{do19},  \cite{chen19}). %In particular, joint SLU models based on BERT have been shown to give strong SOTA results \cite{chen19}. However, 
%while there is rich body of work on SLU, 

Academic research on SLU has mostly focused on improving overall accuracy on datasets with a static distribution and a fixed set of intents. By contrast, in real-world SLU applications, new intents are continuously added over time, yielding dynamically changing and highly imbalanced data distributions.
%, where initial support for new intents needs to be bootstrapped without any live data samples. 
This issue, which is known as \textit{data imbalance} in machine learning \cite{survey}, usually leads to poor predictive performance on minority classes. However, for an SLU application, it is important to support all intents, because the set of supported functionalities is announced to the customers, and there is no direct correspondence between the frequency with which a certain functionality is used and its importance. For example, an utterance "where is my phone?" is likely rarely used, but may be still important to the customer. Notably though, in Academic SLU research the problem of low performance for low-frequency intents is typically hidden, as overall IC accuracy is measured, which is governed by performance of the head intents, as the percentage of test samples for tail classes is low.
%though they may be important, depending on the task. For example, in disease detection, detecting a low-frequency disease class with maximum accuracy may be very important, and in image processing tasks one may be interested in achieving similarly high performance for each class, including low-frequency ones. 
%with several classes having no or few data samples. 

In industry research, recently the data sparseness problem for new classes has been explored. Corresponding approaches investigated in particular data augmentation with synthetic data, such as machine translated (MT) data \cite{gaspers2018} or data collected via 
%semi-supervised learning and/or 
paraphrase generation techniques \cite{ssl-para}. However, typically, the focus has been on how synthetic data can be generated, and sparse classes were simply balanced by adding synthetic data without taking overall data distributions into account.%To the best of our knowledge, the more general problem of handling data imbalance for SLU has not yet been explored in a systematic manner.  

To the best of our knowledge, this paper provides the \textit{first} systematic study on handling data imbalance for SLU. In particular, we study the impact of data imbalance and evaluate different data balancing approaches for SLU.
%In general, data imbalance is a known problem in ML, posing s a challenge for predictive modeling as almost all classification algorithms are designed around the data balance assumption which may lead to poor predictive performance for minority classes. Unfortunately, the problem has not been studied sufficiently in the literature for modern DNN-based models yet, and several findings are not consistent across datasets and tasks \cite{survey}, indicating that different methods may be beneficial for different tasks and goals. For example, in certain image processing tasks one may be interested in achieving similarly high performance for each class, while in disease detection the focus may be on detecting a low-frequency disease class with maximum accuracy. 
Our task has the following \textit{specific challenges}: 
%i) we explore a joint model for two subtasks and thus need to deal with two data distributions,
i) the class distribution used by customers during application is unknown during model training time, because it is evolving over time as new features are added and user behaviour changes, and ii) our aim is to boost performance of low-frequency intents, while not decreasing accuracy for the head intents. The latter is important, as the head intents are frequently used by customers, and thus 
%they are used to their comparatively high performance. 
a decrease in accuracy could risk a drop in customer satisfaction with the device. 
%Thus, deploying performance is typically not acceptable.
%Towards this goal

While there is a large body of work addressing the data imbalance problem, it has not been studied sufficiently yet for modern DNN-based models and several findings are not consistent across tasks \cite{survey}, indicating that different methods may be beneficial for different tasks. In addition, common data balancing methods like random over-sampling or weighted loss functions (e.g. \cite{Chawla_2002}, \cite{Guo2017LearningFC}, \cite{survey}) may cause over-fitting, 
%The most common methods for boosting performance of low-frequency classes include weighted loss functions increasing the impact of minority classes during optimization and random or synthetic over-sampling of low frequency classes (e.g. \cite{Chawla_2002}, \cite{Guo2017LearningFC}, \cite{survey}). However, these approaches may cause over-fitting, 
potentially leading to a performance drop of the head intents on unseen data during application.

To overcome the over-fitting problem, inspired by Zhang et al. (2019) \cite{zhang2019balance}, we use a multi-task framework including a primary task and an auxiliary task sharing a common feature extraction component. Although the two tasks are trained alternatively in a multi-task setting, only the primary decoders are used during the inference phase. Thus, if we add extra information to the auxiliary task, it will be indirectly injected to the primary task through the feature extraction. Intuitively, this can help us to prevent the primary model from over-fitting.
% and to reduce the negative impact of low quality information on the primary model.
%While our primary task is a standard SLU model, 
In this work, we apply two data imbalance handling techniques, i.e. training with a class-balanced batch generator and data augmentation with synthetic data to the auxiliary task.

We present empirical results on a real-world imbalanced SLU dataset. In particular, we compare our proposed approach with applying standard techniques including over-sampling, data augmentation with synthetic data, a re-weighting scheme with a DNN loss, and a class-balanced batch generator, directly on the primary task. Our results indicate that: 
\begin{enumerate}
    \setlength{\parskip}{-3pt}
    \item our proposed model can boost performance on low frequency intents significantly while avoiding a performance decrease on the head intents which is a potential issue faced by common data re-sampling and re-weighting methods, 
    \item synthetic data are useful for intent bootstrapping,
    %new intents, 
    but
    \item once a certain amount of realistic intent data becomes available, using synthetic data in the auxiliary task only yields better performance than adding them to the primary task training data, and
    \item in a joint training scenario, balancing the intent distribution individually improves both IC and SF performance.
\end{enumerate}
% !TEX root = ../coling2020.tex 

\section{Related Work}
We are not aware of previous work addressing the data imbalance problem in a systematic manner for SLU. However, there is a large body of work addressing this topic in other fields, particularly in image processing (e.g. \cite{Guo2017LearningFC}, \cite{survey}). The most common  approaches to data balancing include different re-sampling and re-weighting strategies. In random over-sampling, samples for the minority classes are duplicated \cite{Guo2017LearningFC}. While this has been shown to be quite effective, it increases model training times and has been found to cause over-fitting \cite{overfitting}. In addition, synthetic over-sampling of minority classes (SMOTE \cite{Chawla_2002}) as well as random under-sampling of majority classes \cite{rus} has been explored. %As in the latter a certain amount of head class samples is discarded, important information may be lost, potentially decreasing model performance on the head classes.
Weighted loss functions can be applied to change the impact of (certain) classes during optimization.
%, e.g. to increase the impact of minority classes. 
A common strategy for re-weighting a loss function  is using the (smoothed) inverse of class frequencies (e.g. \cite{huang2016lmle}, \cite{mahajan2018exploring}). 
Moreover, leveraging information about the class distribution on a reference dataset has been explored, for instance, dynamic sampling \cite{dynsam} and a label-distribution-aware margin loss  \cite{NIPS2019_8435} have been explored. However, a challenge in our task is that the class distribution during application is unknown during model training and hence cannot by leveraged. %In particular new intents need to be modeled before they appear in the application's data.

Another related line of research explores boosting performance of low-frequency classes via knowledge transfer from head to tail classes. (e.g. \cite{Ouyang_2016}, \cite{NIPS2017_7278}).

Zhang et al. (2019) explore data balancing for image processing tasks by making use of an auxiliary task which combines a class-balanced and a random batch generator \cite{zhang2019balance}. We have adapted this idea for our SLU task, e.g. towards handling joint tasks, to use different batch generators for primary and auxiliary tasks, and we additionally study the integration of synthetic data. 

Different methods have been proposed for generating synthetic data to boost performance of new languages or features in SLU via data augmentation (e.g. \cite{gaspers2018}, \cite{ssl-para}). While we include synthetic data into our study, our focus is not on data generation.

% !TEX root = ../coling2020.tex                     
\section{Method}
%Current state-of-the-art approaches to SLU \cite{chen19} model IC and SF jointly based on BERT \cite{mbert}. We follow the state-of-the-art in that we base our model upon BERT as well in order to get a strong baseline, however, we extend it towards handling data imbalance, specifically by including an auxiliary task. 
%A potential obstacle in applying data balancing methods to a real-world problem is over-fitting. Since the real data distribution is often unknown at training time, adjusting the training according to either a balanced distribution or a development distribution to improve low-frequency classes, potentially leads to a performance drop on the head intents at application time. 

%To overcome over-fitting, 
For the SLU task, we use a joint model for IC and SF based on BERT \cite{mbert}. We develop a multi-task framework consisting of two of these SLU tasks i.e. a primary and an auxiliary SLU task sharing a common feature extraction. While the two tasks are trained jointly, only the primary task decoders are used for the inference phase. The \textit{auxiliary task} is trained using \textit{a class-balanced batch generator} (CBG) and (optionally) \textit{data augmentation with synthetic data}. %Intuitively, this ``indirect" injection of extra information can help us not only to prevent the primary model from over-fitting but also to reduce the negative impact of low quality synthetic data on the primary model.

%Thus, instead of applying data balancing techniques directly for the primary task, which is the common approach, we only use them in the auxiliary task. In particular, the auxiliary task in our model makes use of a class-balanced sampled batch generator, and possibly includes synthetic data. With this approach, we hope that useful knowledge from the auxiliary task can strengthen the feature extraction for dealing with data imbalance, while preventing the primary model from overfitting. 

\subsection{SLU Model}
\begin{figure}%[!ht]
\centering
 \includegraphics[scale=0.2]{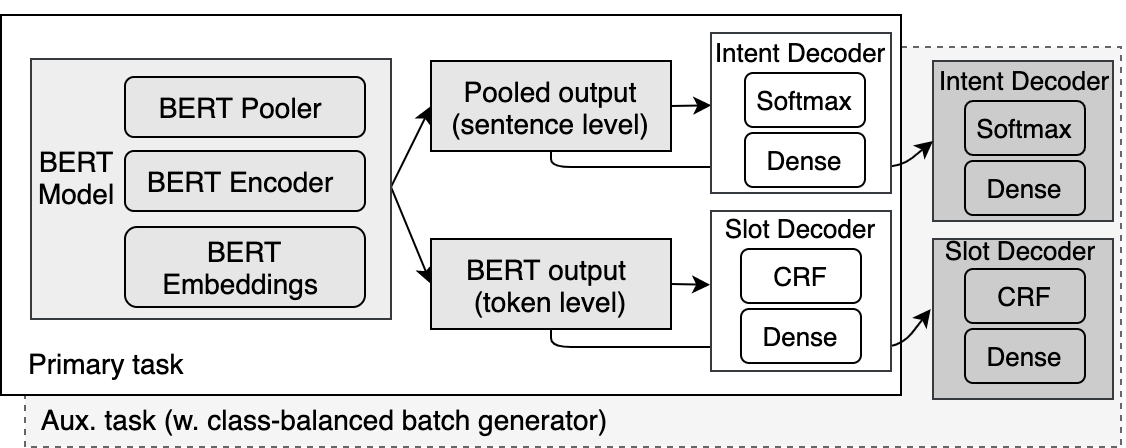} 
  \caption{A multi-task model for handling data imbalance with a class balanced batch generator.}
  \label{fig:model}
  \vspace{-0.5cm}
\end{figure}

Figure \ref{fig:model} shows our multi-task model for handling data imbalance in SLU. Our model deals with two tasks: 
\begin{itemize}
    \item Primary task: Standard SLU, which is a joint model of IC and SF using a regular random batch generator.
    \item Auxiliary task: SLU using a class balanced batch generator, which has the same model architecture as the primary task, but is trained with a special batch generator to assure class balance in each training batch. Moreover, synthetic data may be used during training to balance low-frequency classes with synthetic samples.
\end{itemize}

The SLU model consists of a BERT \cite{mbert} encoder, an intent decoder and a slot decoder. As shown in Figure \ref{fig:model}, the BERT encoder's outputs at sentence level and token level are used as inputs for the intent and slot decoders, respectively. The intent decoder is a standard feed-forward network including two standard dense layers and a softmax layer on top. The slot decoder uses a CRF layer on top of two dense layers to leverage the sequential information of slot labels. The two losses of IC and SF are optimized jointly with balanced weights (1.0:1.0).

To perform multi-task training, we alternate through tasks using a ratio of 1.0:1.0 between primary task batches and auxiliary task batches, respectively, during the training process. The training process is completed when both of the tasks are done.

\subsection{Class-Balanced Batch Generator For A Joint Model}
\begin{algorithm}%[H]
\SetAlgoLined
\KwInput{$\b D$ - training data set, $\b C$ - set of classes (labels), and $b$ - batch size}
 size=0\; size\_per\_class=b/$|\b C|$;
 
 \While{$size \leq |\b D|$}{
  new\_batch = $\emptyset$
  
  \For{$c$ in $\b C$}{
    $\b S$ = \{ size\_per\_class instances of class $c$ selected randomly from $\b D$ \}
    
    new\_batch = $\b S$  $\cup$ new\_batch
    }
    
  size=size+$|$new\_batch$|$
  
  train model using new\_batch
 }
 \caption{CBG}
   \label{alg:cbs}
\end{algorithm}
Algorithm \ref{alg:cbs} shows our class-balanced batch generator. Each batch contains the same number of instances per class. To avoid training time explosion on large-scale datasets, we stop an epoch when the total number of the generated instances in the epoch exceeds the original training size.

%Handling data imbalance in a joint model is a \textit{challenge} as 
We have to deal with two data distributions due to having two sub-tasks. Since slot labels are often shared across several intents, we assume that there is a strong correlation between the distributions so that when the IC distribution is balanced, the slot distribution is balanced accordingly. Therefore, our class-balanced batch generator is performed \textit{on the IC distribution only}.
\subsection{Data Augmentation With Synthetic Data}
Due to language expansion of voice-controlled devices, a large amount of annotated SLU data may be available in different languages, in particular in English. These data can be leveraged for SLU model development in another language (e.g. \cite{gaspers2018},  \cite{do19}, \cite{johnson-etal-2019-cross}). In this work, following Gaspers et al. (2018) \cite{gaspers2018}, we automatically translate English source data into our target language, and use the translated data as extra training data in the auxiliary task. Thus, in contrast to previous work, our approach supports utilizing synthetic data for re-balancing rather than just directly for model training. This can help us to avoid the negative impact of low quality synthetic data on the primary task.

% !TEX root = ../coling2020.tex     
\section{Experiments}
%In this section, we first describe the utilized Alexa SLU datasets and subsequently the experimental settings.
In this section, we describe our experiments comparing different data balancing techniques on a real-world German SLU dataset. 
\subsection{Datasets}
\begin{table}%[ht]
\centering%\small
\begin{tabular}{l|l|l|l|l}
\hline
\textbf{Dataset} & \textbf{Snooze} & \textbf{Browse} & \textbf{Cancel} & \textbf{Head I.}\\
\hline
\hline
Train & 40 & 16& --& 166,493\\
Val. & 10 & 4 & -- & 20,846\\
Test & 449 & 634 & 623& 20,710 \\
NMT data & 20,285 & 7,855& 2,657 & --\\
\hline
\end{tabular}
\caption{Number of samples for three simulated low-frequency intents and the remaining seven intents (head intents).}
% SnoozeNotificationIntent, BrowseReminderIntent. and CancelReminderIntent. have 50, 20 and 0 samples in training and validation datasets, respectively.}
\label{tbl:data}
%\vspace{-0.3cm}
\vspace{-1.0cm}
\end{table}
\paragraph*{Realistic SLU data} We extracted a random data sample
from a commercial German SLU system; the data are representative of requests to voice-controlled devices and were manually annotated with intent and slot labels (see Table \ref{tbl:data}). The sample is from the \emph{Notifications} domain and comprises 351.066 samples. We split the dataset into 80\% train, 10\% validation and 10\% test data. The data are highly imbalanced for intent classes with a long-tail distribution, where the smallest classes comprise a single sample only. Due to the small amount of test instances for low-frequency intents, we cannot reliably evaluate performance for them. Therefore, we created an artificial set up to study intent bootstrapping in a systematic way. In particular, we first removed data of all intents having fewer than 1,000 samples, leaving 10 classes. This ensures that at least 100 test samples are available per intent, which are needed to measure performance reliably in a large-scale set up. We selected the three intents with the lowest frequency from the remaining intents to simulate intent bootstrapping and filtered out all of their samples from the train and validation datasets. To simulate the growing data amounts per class in a developing system, where new features are added over time, we randomly sampled a different amount for each of the three intents and re-added 80\% and 20\% to the filtered train and validation datasets, respectively. We added 0, 20 and 50 samples for \emph{CancelReminderIntent}, \emph{BrowseReminderIntent} and \emph{SnoozeNotificationIntent}, respectively. 
%This is the \textit{main} data set for our SLU experiments.
\paragraph*{Machine-translated data} We translated data from an English NLU system into German using a transformer-based neural machine translation (NMT) system trained with Sockeye \cite{Sockeye:17}. The NMT system was trained on 4,000 segments of internal data as well as 28,733,606 segments of publicly available data; slot labels were projected from the English source utterances to the German translations using the alignment model \emph{fast\_align} \cite{fastalign}. The number of NMT-generated samples is 2,657, 7,855 and 20,285 for \emph{CancelReminderIntent}, \emph{BrowseReminderIntent} and \emph{SnoozeNotificationIntent}, respectively. This data set is used for \textit{data augmentation} of low-frequency classes in our experiments.

%\textbf{Publicly available SLU data.}
%The ATIS dataset \cite{conf/slt/TurHH10} is a small-scale dataset, which has been widely used in SLU research. It consists of audio recordings and corresponding annotated transcriptions in English of people making flight reservations. We use the version provided by \newcite{N18-2118}, in which the training, validation and test sets contain 4.478, 500 and 893  utterances, respectively. The data are imbalanced, with the smallest intent classes comprising a single sample only. %The word vocabulary of the dataset is of size 778.

\subsection{Models}
We train and evaluate the following models on our SLU data:
\begin{itemize}
    \setlength{\parskip}{-3pt}
    \item \emph{Baseline}: The standard model without any data imbalance handling technique; this baseline is obtained by training the primary task individually on SLU training data.
    \item \emph{Over-sampling}: Common random over-sampling method; for each intent class, SLU training data are up-sampled to the number of samples in the head class. The primary task is then trained individually on the up-sampled data.
    \item \emph{Balanced-loss}: Common re-weighting method with DNN loss; The cross entropy loss of IC is balanced by using class frequencies. The model is obtained by training the primary task individually on SLU training data.
    \item \emph{CBG}: The primary task is trained individually on SLU training data using the class-balanced batch generator.
    \item \emph{Mul.-CBG}: Proposed method \textit{without} data augmentation; an auxiliary task with class-balanced batch generator (without using machine translated data) is applied.
    \item \emph{Data-aug.}: Common training data augmentation method; SLU training data are augmented with the synthetic NMT data for low-frequency classes. The primary task model is then trained individually on the augmented data.
    \item \emph{Data-aug.+Over-sampl.}: Combination of the above \emph{data-aug.} and \emph{over-sampling} approaches. SLU training data are first augmented by adding all of the available synthetic NMT data for low-frequency classes and subsequently up-sampled as in \emph{over-sampling}. The model is then obtained by training the primary task individually on the up-sampled and augmented training set.
    \item \emph{Data-aug.+Balanced-loss}: Combination of the above \emph{data-augmentation} and \emph{balanced-loss} approaches. SLU training data is augmented by adding the available NMT data for low-frequency classes. The model is then obtained by training the primary task individually on augmented training data with balanced loss.
    \item \emph{Data-aug.+CBG}: Combination of the above \emph{data-augmentation} and \emph{CBG} approaches. SLU training data is augmented with the available NMT data. The primary task model is then trained individually on the augmented SLU training set using a class-balanced batch generator.
    \item \emph{Mul.-CBG+Data-aug.}: Proposed method \textit{with} data augmentation; the auxiliary task with class-balanced batch generator is applied on the NMT-augmented dataset.
\end{itemize}
%While \emph{Ours} and  \emph{Ours+Data-aug.} are trained using both of primary and auxiliary tasks, the other models are trained with only the primary task in our model.
\subsection{Settings}
In our experiments, we use pre-trained multilingual BERT \cite{mbert} (size 768), and max-pooling for sentence representation. Each of our decoders has 2 dense layers of size 768 with gelu activation. The dropout values used in IC and SF decoders are 0.5 and 0.2, respectively. For optimization, we use Adam optimizer with learning rate 0.1 and a Noam learning rate scheduler. We trained our model 
%for 10 epochs 
with batch size of 64.

Following Gaspers et al. (2018) \cite{gaspers2018}, we use a semantic error rate, which measures IC and SF jointly and is defined as follows:
\begin{equation}
\textrm{\emph{SemER}} = \frac{\#(\textrm{slot+intent errors})}{\#\textrm{slots in reference + 1}}\end{equation}

\section{Results}
%In the following, we first conduct a systematic study regarding data balancing for large-scale SLU on our Alexa feature bootstrapping set up. Subsequently, we explore on the ATIS benchmark datset whether using an auxiliary task with class-balanced batch geSFator can improve upon a SOTA SLU model.
%\subsection{Feature bootstrapping on internal data}

Table \ref{tbl:fb_results} shows the performances of our experimental models on the head intents, and on each of the three low-frequency intents.%, i.e.  \emph{SnoozeNotificationIntent}, \emph{BrowseReminderIntent} and \emph{CancelReminderIntent}. %Note that among the three low-frequency intents, \emph{CancelReminderIntent} is special in that it represents a completely new intent, where no live data are available yet, and hence there are no regular training data for it in our set up.
\begin{table}[t]
%\small
\resizebox{1.05\columnwidth}{!}{%
\begin{tabular}{l|l|l|l|l}
\hline
\textbf{Method} & \textbf{Snooze} & \textbf{Browse} & \textbf{Cancel} & \textbf{Head I.}\\
\hline
\hline
%\emph{Baseline}  & 22.56 & 64.43 & 55.58 & 4.55\\

\emph{Over-sampling} &-30.63&
-56.03&
-16.77&
+7.03\\
\emph{Balanced-loss} & -32.45&
-40.37&
-16.77&
+11.43\\
\emph{CBG} & -40.56 &
-56.77&
-8.10&
+8.57 \\ \hline
\emph{\textbf{Mul.-CBG} (ours)}  &-3.59&
-45.77&
-9.93&
-1.10 \\ \hline \hline
\emph{Data-aug.} & +236.97&
-32.66&
-43.83&
-0.22\\
% CBS 6
\emph{Data-aug.+Over-sampl.} & -33.33&
-37.16&
-44.98&
+3.08 \\
\emph{Data-aug.+Balanced-loss} & -57.67 &
-32.98&
-49.42&
+3.96\\
\emph{Data-aug.+CBG} & -64.85&
-37.25&
-45.09&
+19.56\\
\hline
\emph{\textbf{Mul.-CBG+Data-aug.} (ours)} & -29.74 &
-34.21&
-2.63&
-0.44\\
%\emph{base},  & mBERT on \emph{base} +  & \multirow{2}{*}{15.85} & \multirow{2}{*}{42.39} & \multirow{2}{*}{54.12} & \multirow{2}{*}{4.53}\\
%\emph{base+MT} & CBS on \emph{base+MT} &&&& \\
\hline
\end{tabular}}
\caption{Relative change in semantic error rate (\%) on three low-frequency intents and on the head intents for data balancing methods on real-world SLU data compared to the baseline. A decrease in SemER implies better performance.}. 
% SnoozeNotificationIntent, BrowseReminderIntent. and CancelReminderIntent. have 50, 20 and 0 samples in training and validation datasets, respectively.}
\label{tbl:fb_results}
\vspace{-0.9cm}
\end{table}

\subsection{Performance without data augmentation} Performance for all of the low-frequency intents can be improved just by balancing the regular training data. However, for \emph{CBG}, random \emph{over-sampling} and \emph{balanced-loss} methods, the boost in performance comes at the cost of a relative increase in \emph{SemER} for the head intents of at least 7.03\%. As head intents are much more frequently used by customers, and the customers are used to their comparatively high performance already, such a decrease in performance is usually not acceptable. By contrast, our approach with the class-balanced batch generator boosts performance for all low-frequency features, yielding up to 45.77\% relative reduction in \emph{SemER} per intent, without over-fitting on the low-frequency classes. Thus, our results suggest that leveraging data balancing techniques in the auxiliary rather than the main task is beneficial. This may be the case, because the model keeps the access to important information about the intent distribution, which is lost when data balancing is applied in the primary task.   
\subsection{Performance with data augmentation} As expected, without using any other techniques, performance is improved when simply adding NMT data (\emph{data-aug.}) to \emph{CancelReminderIntent}, which didn't have any intent samples beforehand. However, for \emph{SnoozeNotificationIntent}, which already had 50 intent samples, simply adding NMT data (\emph{data-aug.}) yields 236.97\% increase in \emph{SemER}. This highlights the fact that in an evolving system one needs to be careful with maintaining synthetic data. They can be very useful to bootstrap new features when no or few realistic feature data are available. However, as they are typically of comparatively low quality, they can decrease performance when they are kept while more and more realistic data become available. While random \emph{over-sampling}, \emph{CBG} and \emph{balanced-loss} can mitigate this problem and improve performance on low-frequency classes, this again comes at the cost of a decrease in performance for the head intents (cf. \emph{data-aug.+over-sampl.}, \emph{data-aug.+CBG} and \emph{data-aug.+balanced-loss}).
\emph{Multi-task-CBG+Data-aug.} indicates performance when NMT data are not added to the primary task training data, but used in the auxiliary task class-balanced batch generator during training only. For \emph{SnoozeNotificationIntent} and \emph{BrowseReminderIntent}, which already have a small number of samples in the realistic training data, performance is improved compared to adding NMT data to the primary task training data without overfitting on the head classes. For \emph{SnoozeNotificationIntent} performance for \emph{Mul.-CBG+Data-aug.} is better than both \emph{Mul.-CBG} without using data augmentation and \emph{data-aug}. This suggests that once a certain amount of realistic intent data becomes available, it's beneficial to leverage NMT data only with \emph{Mul.-CBG+Data-aug}. However, for intents without realistic training data, standard NMT data augmentation is preferable, as the intent cannot be recognized otherwise. Note that this is the case for \emph{CancelReminderIntent} in this experiment. 
\subsection{Does balancing the intent distribution help slot filling?} Interestingly, the intent-based data balancing methods also improve SF performance, as indicated by the results for \emph{CancelReminderIntent} by several methods without including data augmentation, in particular \emph{Mul.-CBG}. Recall that there are no samples for this intent in the regular training data. Hence, the intent cannot be recognized, adding one error to \emph{SemER} for each test utterance evaluation independent of the method (not applying data augmentation). In fact, the improvements for \emph{CancelReminderIntent} are solely resulting from improved SF performance. We attribute this improvement to the fact that slots are shared across several intents and may benefit from intent-based data balancing, as we are using a multi-task model for IC and SF.  
\subsection{Performance on benchmark data}
Unlike previous approaches to SLU, our method is focused on improving low-frequency classes which have typically very few samples in small-scale benchmark datasets and thus improvements in the overall standard SLU error metrics IC accuracy and slot F1 might not be expected. However, to investigate whether using an auxiliary task with a class-balanced batch generator improves upon a SOTA SLU model on a benchmark dataset we evaluated our approach on the small-scale ATIS dataset \cite{conf/slt/TurHH10} using the version provided by \cite{N18-2118}. In particular, we trained the baseline and \emph{Mul.-CBG} models, and we found that \emph{Mul.-CBG} improves over the baseline on IC accuracy and slot F1 from 97.4\% to 97.5\% and from 95.7\% to 96.2\%, respectively. This indicates that our method can even increase overall performance on small-scale benchmark data, and that our model yields comparable performance to state-of-the-art SLU systems \cite{chen19, do19, N18-2118}. In addition, the results provide further evidence that balancing the intent distribution also improves SF, i.e. from 95.7\% to 96.2\% in F1, where the latter even slightly outperforms the previously best reported results of 96.1\% for SF on ATIS \cite{chen19}. 
\section{Conclusion}
We presented a study comparing different techniques for handling data imbalance for DNN-based SLU. Aiming to boost performance for low-frequency intents, we proposed
a multi-task model for SLU in which we make use of an auxiliary task to deal with data imbalance. 
%In particular, the auxiliary task is trained using a class-balanced batch generator, which may include synthetic data. 
Our results on a real-world SLU dataset indicate that: i) in contrast to common data re-sampling and re-weighting methods, our method can boost performance on low frequency intents significantly without decreasing performance of the head intents, ii) synthetic data are beneficial for bootstrapping new intents when realistic intent data are not available, but iii) once a certain amount of realistic intent data becomes available, using synthetic data in the auxiliary task only yields better performance than adding them to the primary task, and iv) in a joint training scenario, balancing intent distribution individually improves not only intent classification but also slot filling performance. Overall, our method achieved relative error rate reductions of up to 45.77\% for low-frequency intents.  
%relative reduction in SemER per intent, without over-fitting on the low-frequency classes

%, and iii) on publicly available benchmark data, our method can improve upon the state-of-the-art models. 

\bibliographystyle{IEEEtran}

\bibliography{is}

% \begin{thebibliography}{9}
% \bibitem[1]{Davis80-COP}
%   S.\ B.\ Davis and P.\ Mermelstein,
%   ``Comparison of parametric representation for monosyllabic word recognition in continuously spoken sentences,''
%   \textit{IEEE Transactions on Acoustics, Speech and Signal Processing}, vol.~28, no.~4, pp.~357--366, 1980.
% \bibitem[2]{Rabiner89-ATO}
%   L.\ R.\ Rabiner,
%   ``A tutorial on hidden Markov models and selected applications in speech recognition,''
%   \textit{Proceedings of the IEEE}, vol.~77, no.~2, pp.~257-286, 1989.
% \bibitem[3]{Hastie09-TEO}
%   T.\ Hastie, R.\ Tibshirani, and J.\ Friedman,
%   \textit{The Elements of Statistical Learning -- Data Mining, Inference, and Prediction}.
%   New York: Springer, 2009.
% \bibitem[4]{YourName17-XXX}
%   F.\ Lastname1, F.\ Lastname2, and F.\ Lastname3,
%   ``Title of your INTERSPEECH 2020 publication,''
%   in \textit{Interspeech 2020 -- 20\textsuperscript{th} Annual Conference of the International Speech Communication Association, September 15-19, Graz, Austria, Proceedings, Proceedings}, 2020, pp.~100--104.
% \end{thebibliography}

\end{document}